# Metal-Insulator Transition and Ferromagnetism in the Electron Doped Layered Manganites La$_{2.3-x}$Y$_x$Ca$_{0.7}$Mn$_2$O$_7$ (x=0,0.3,0.5)

P. Raychaudhuri[*], C. Mitra, A. Paramekanti, R. Pinto, A. K. Nigam, S. K. Dhar

*Tata Institute of Fundamental Research,
Homi Bhaba Road, Colaba,
Mumbai-400005
India.*

***Abstract:*** Bulk samples of La$_{2.3-x}$Y$_x$Ca$_{0.7}$Mn$_2$O$_7$, x=0,0.3,0.5, with layered perovskite structure have been synthesized and investigated with respect to their electrical, electronic and magnetic properties. It is found that La$_{1.8}$Y$_{0.5}$Ca$_{0.7}$Mn$_2$O$_7$ has tetragonal structure and is a metallic ferromagnet with a magnetic transition temperature of 170 K. The compound shows metallic behavior below 140 K and has a large magnetoresistance (MR) $|\Delta\rho/\rho_0|$ ~94% at 100 K at 34 kOe. For x=0 and 0.3 the structure is monoclinic with a suppression of metallicity. For x=0 the material is an ferromagnetic insulator. We observed a large increase in the coefficient of the linear term in specific heat with decreasing x. To our knowledge this is the first report of an electron doped manganite showing metal-insulator transition and ferromagnetism.

[*]e-mail:pratap@tifrc3.tifr.res.in

Hole doped rare-earth manganese oxides have attracted considerable attention in recent times because of their magnetotransport phenomenon arising from spin-charge coupling. Hole doping is achieved by partially replacing the trivalent rare-earth ion by a bivalent ion like Sr, Ca or Ba by which the manganese ion goes in a mixed $Mn^{3+}/Mn^{4+}$ valence state. Materials with the form $R_{1-x}M_xMnO_3$ (R is a rare-earth ion whereas M is a bivalent cation) with 3-dimensional Mn-O-Mn bonds like, for example, in $La_{1-x}Ca_xMnO_3$ and $La_{1-x}Sr_xMnO_3$ are known for a long time to be paramagnetic insulators at high temperatures and ferromagnetic metals at low temperatures (for x>0.2). In contrast $R_{2-2x}M_{1+2x}Mn_2O_7$ compounds with layered perovskite structure have 2-dimensional Mn-O-Mn network in the a-b plane. Studies on $La_{2-2x}Sr_{1+2x}Mn_2O_7$ have shown that these materials show strong anisotropy in the a-b plane (in plane) and along the c axis (out of plane) [1]. Ferromagnetism and metal-insulator transition in this class of compounds have been reported earlier in the hole doped $La_{2-2x}Sr_{1+2x}Mn_2O_7$ [1,2] and $La_{2-2x}Ca_{1+2x}Mn_2O_7$ (x>0.4) [3]. In $Sr_{2-x}Nd_{1+x}Mn_2O_7$ [4] the resistivity shows a complex behavior but the material does not show any ferromagnetic transition. In all these compounds the magnetic and electrical properties are governed by Zener double exchange mechanism based on the mixed $Mn^{3+}/Mn^{+4}$ valence states [5].

In this letter we report the magnetic, transport and electronic properties of the electron-doped manganites $La_{2.3-x}Y_xCa_{0.7}Mn_2O_7$ (x=0, 0.3, 0.5). Unlike the compounds reported earlier here the R/M ratio is such that manganese goes in a mixed $Mn^{2+}/Mn^{3+}$ valence state. The double exchange operates here in the $Mn^{+2}$-O-$Mn^{+3}$ bonds giving rise to ferromagnetism and metal-insulator transition. The compound with x=0.5 is a metal below

140 K and has a ferromagnetic transition temperature of 170 K. The material shows large negative magnetoresistance ($|\Delta\rho/\rho_0|$ ~94% at 3.8 Tesla at 100 K).

Bulk ceramic samples were prepared through the conventional solid state reaction route starting from $La_2O_3$, $Y_2O_3$, $CaCO_3$ and $MnO_2$. Stoichiometric amounts of the starting oxides and carbonates were mixed, ground and calcined in air for 18 hrs at $900^0C$. The reacted powder was then reground, pelletised, sintered for 15 hrs at $1450^0C$ in oxygen flow, cooled down to $1050^0C$ at $10^0C$/min kept for 10 hrs in oxygen flow, and cooled to room temperature at $10^0C$/min. The samples were characterized through X-ray diffraction (XRD) and energy dispersive X-ray microanalysis (EDX), and scanning electron microscope (SEM). The cell constants were calculated using the XLAT software for the tetragonal structure. The composition was found to be nearly identical to the starting composition within the accuracy of 3% of EDX. Electrical resistance and magnetoresistance was measured using the conventional 4-probe technique. The magnetoresistance (MR) was defined as $\Delta\rho/\rho_0=(\rho(H=0)-\rho(H))/\rho(H=0)$. The magnetization was measured either using a Quantum Design SQUID magnetometer or by the Faraday method. Specific heat of these samples was measured using a home made semi adiabatic, heat pulse-type setup calibrated with copper standard with an accuracy of 4%.

Figure 1 shows the representative XRD patterns of $La_{2.3-x}Y_xCa_{0.7}Mn_2O_7$ (x=0 and 0.5). All the peaks for the composition x=0.5 are indexed with the tetragonal structure with lattice constants a=3.874 Å and c=19.218 Å. For x=0 and 0.3 the crystal structure becomes monoclinic. For x=0.3 the lattice constants are a=3.881 Å, b=19.285 Å, c=4.003 Å and $\beta=93.16^0$ while for x=0 the lattice parameters are a=3.890 Å, b=19.336 Å, c=4.008

Å and $\beta=93.40^0$. No impurity phase was detected from XRD. It appears that the tetragonal structure is obtained when the average ionic radius at the rare-earth/alkaline-earth metal site is same as that of $La_{1.5}Ca_{1.5}Mn_2O_7$ which has a well known tetragonal structure [3]. In this structure-type the Mn-O-Mn bonds are separated along the c-axis by R(Ca)O layers. This causes the double exchange mechanism to be stronger in the a-b plane than out of plane, which makes the system quasi two-dimensional. This has interesting effects in the electronic and magnetic properties of the material. Figure 2 shows the SEM photograph of the sample with x=0.5 showing the grain structure with average grain size of 3μm.

The resistivity ρ as a function of temperature for $La_{2.3-x}Y_xCa_{0.7}Mn_2O_7$ (x=0,0.3,0.5) and MR as a function of field (up to 3.8 Tesla) for x=0.5 at various temperatures up to 100 K is shown in figure 3a and 3b respectively. The resistivity as a function of temperature (R-T) shows a peak at ($T_p$) 140 K for x=0.5 and at 107 K for x=0.3. We do not observe any metallicity for the sample with x=0 though there is a slight rounding off in the R-T curve at low temperature. The residual resistance of the sample increases by almost seven orders of magnitude with change of composition from x=0.5 to x=0. This is in sharp contrast with the perovskite manganites where doping of smaller rare-earth like yttrium at the cation site decreases the metallicity in the sample[6]. The way the cation size affects the metallicity of the sample depends on the the way the Mn-O-Mn bond angle changes with change in ionic size. In the case of perovskite manganites with substitution of small rare-earth the bond angle deviates from the ideal $180^0$, which in turn reduces the hopping integral $t_{ij}$ of the $e_g$ electrons between two neighboring sites. It is not yet very well known how the Mn-O-Mn bond angle gets modified in the case of layered perovskites structure, where we have two types of Mn-O-Mn bonds, one between two manganese between the bilayers and the other between two manganese in the same layer. However, we observe that the highest metallic behavior is obtained when the system goes to the tetragonal phase. The behavior above the peak temperature is semiconducting and metallic below. We observe that the material with x=0.5 has a MR~60% at 4.2 K and 93.8% at 100 K. Figure 4a depicts the zero field cooled magnetization plots measured in applied fields of 1500 Oe and 84 Oe (inset) respectively. All the samples have a magnetic transition temperature ($T_c$) in the range of 160 K to 170 K though the metallicity, for the composition with x=0 is completely suppressed. The magnetization in Bohr magneton per manganese ion as a function of field up to 4 Tesla at 4.5 K for $La_{1.8}Y_{0.5}Ca_{0.7}Mn_2O_7$ is shown in figure 4b. The saturation value 4.33$\mu_B$ is close to the value deduced from the formula for this $Mn^{+2}/Mn^{+3}$ ratio ( 4.15$\mu_B$ ). The apparent lack of correlation between $T_c$ and the metal insulator transition temperature has been earlier observed in these layered perovskite compounds [3]. Asano et al. have suggested that there are two different ferromagnetic coupling possibly from the anisotropic double exchange interaction [3,7]. This might give rise to the large deviation between $T_c$ and $T_p$.

Figure 5 shows the specific heat (C) as a function of temperature (T) for x=0,0.3 and x=0.5. The C versus T curve was fitted to an expression of the type $C=\gamma T+\beta T^3$ in the temperature range 2.5 K to 10 K. The values of $\beta$ and $\gamma$ were calculated from the slope and the intercept of the fitted C/T versus $T^2$ curve at T→0 respectively (inset). The Debye temperature ($\theta_D$) calculated from $\beta$, was found to be 329.5 K for x=0, 365.9 K for x=0.3 and 379.9 K for x=0.5. The smaller value of $\theta_D$ for smaller x is consistent with the fact that compositions with smaller x have larger average atomic mass at the rare-earth/alkaline-earth metal site. The value of $\gamma$ is 14.0 mJ/mole-$K^2$ for x=0.5 and 20.0

mJ/mole-$K^2$ for x=0.3 and 41.5 mJ/mole-$K^2$ for x=0. The γ values observed here are much larger than those observed for $ABO_3$ type manganites which range from 4.4 mJ/mole-$K^2$ to 6.1 mJ/mole-$K^2$ for $La_{0.67}Ba_{0.33}MnO_3$ and 7.8mJ/mole-$K^2$ for $La_{0.8}Ca_{0.2}MnO_3$ [8,9]. We believe that the two dimensional character of the system can play an important role in the enhancement of the density of states (DOS) at the Fermi level, which might be responsible for the large γ observed in our compounds. This point is elaborated in the next paragraph. There is, however, an apparent anomaly in the composition with x=0 where γ, which is normally associated with the electronic contribution to the specific heat, is large though the material does not show metallicity down to 15 K. This is possibly due to magnetic phase separation in the system and is explained later.

In the perovskite manganites the five fold degenerate d-orbital of the manganese ion splits into a three fold degenerate $t_{2g}$ and a two fold degenerate $e_g$ orbital in the oxygen octahedra due to the crystal field [9]. In addition, distortion of the oxygen octahedra can split the two fold degenerate $e_g$ level. In the hole doped samples the electrons in the $e_g$ orbital gets delocalised through Zener double exchange via an effective transfer integral t giving rise to conductivity and ferromagnetism. In a scenario where we have strong in plane versus out of plane anisotropy we have $t_x=t_y\gg t_z$. The monoclinic distortion of the unit cell will introduce an additional reduction in the symmetry, namely $t_x \neq t_y$. In our case the manganese are in $Mn^{+2}/Mn^{+3}$ states with an $e_g$ band which is slightly (~7.5%) more than half filled. To understand qualitatively the effect of two dimensionality on the DOS we have calculated the tight binding DOS for two dimensional (2D) and three dimensional (3D) electronic system neglecting the Coulombic repulsion between electrons and assuming some parameters only to illustrate the point made below. The result is shown in figure 6. It is observed that the DOS is greatly enhanced in the 2D case (where $t_x=t_y=t\gg t_z$) compared to 3D (where $t_x=t_y=t_z=t$) near half filling of the band. The 2D DOS for $t_x \neq t_y$ is also shown in figure 6. We see that the splitting of the Van Hove singularity at the half filling of the band in two-dimensional systems can greatly enhance the DOS in the monoclinic structure for certain fillings of the band. This is likely to be the possible reason for the large increase of DOS in x=0.3 compared to x=0.5.

The apparent anomaly in $\gamma$ for the composition with x=0 is possibly due to magnetic phase separation in the system. Neutron diffraction studies in manganite samples have shown since long back that many of the hole doped manganite go into a mixture of ferromagnetic and antiferromagnetic (or canted antiferromagnetic) phase [10,11]. We observe here that the magnetization at low temperatures (fig. 4) is much smaller in the sample with x=0 than for the samples with x=0.3 and 0.5 both at 1500 Oe and at 85 Oe. Thus possibly the sample with x=0 segregates into a mixture of ferromagnetic and antiferromagnetic phases with unconnected ferromagnetic clusters embedded in a antiferromagnetic matrix. In this situation the bulk heat capacity of the sample would include a large electronic contribution to the heat capacity associated with the ferromagnetic clusters which are metallic but electrical behavior of the sample would be insulating due to the lack of a percolating conducting path. Battle et al [12] have recently observed from neutron diffraction studies that the compound $Sr_{2-x}Nd_{1+x}Mn_2O_7$ are naturally biphasic with two very closely related phase of which only one shows long range ferromagnetic ordering. It is not clear whether a similar situation occurs in the present compound. In this context it might be noted that from neutron diffraction studies Perring et al [13] had observed that the paramagnetic phase $La_{1.2}Sr_{1.8}Mn_2O_7$ consists of long lived antiferromagnetic clusters co-existing with ferromagnetic critical fluctuations.

In summary we have successfully synthesized the electron doped layered manganites $La_{2.3-x}Y_xCa_{0.7}Mn_2O_7$, x=0, 0.3, 0.5, showing ferromagnetism and metal insulator transition for x=0.3 and 0.5. The materials display interesting behavior in their magnetic transport and electronic properties. It would be interesting to study the electron doped phase in detail since the manganese $e_g$ band which is responsible for metallicity and ferromagnetism is symmetric with respect to half filling of the band in the parent material (with all $Mn^{+3}$ ions). We believe that apparent lack of correlation between $T_c$ and metal-insulator transition and the enhancement of γ with increasing monoclinic distortion can be understood by considering the effect of anisotropic double exchange in the Mn-O-Mn network.

One of the authors (PR) would like to thank Srikantha Sil for pointing out the role of two-dimensionality in the DOS in these system. The authors would also like to thank S.B. Roy for his help regarding SQUID measurements.

**References:**

1. Y. Morimoto, A. Asamitsu, H. Kuwahara, and Y. Tokura, Nature **380**, 141 (1996)

2. R. Seshadri, C. Martin, M. Hervieu, B. Raveau, and C. N. R. Rao, Chem. Mater. **9,** 270 (1997)

3. H. Asano, J. Hayakawa, and M. Matsui, Appl. Phys. Lett. **68,** 3638 (1996)

4. P. D. Battle, S. J. Blundell, M. A. Green, W. Hayes, M. Honold, A. K. Klehe, N. S. Laskey, J. E. Millburn, L. Murphy, M. J. Rosseinsky, N. A. Samarin, J. Singleton, N. E. Sluchanko, S. P. Sullivan, and J. F. Vente, J. Phys.: Condens. Matter **8**, L427 (1996)

5. C. Zener, Phys. Rev. **82,** 403 (1951)

6. J. Fontcuberta, B. Martinez, A Seffar, S. Pinol, J. L. Garcia-Munoz and X Obrados, Phys. Rev. Lett. **76,** 1122 (1996)

7. H. Asano, J. Hayakawa, and M. Matsui, Appl. Phys. Lett. **70,** 2303 (1997)

8. J. J. Hamilton, E. L. Keatley, H. L. Ju, A. K. Raychaudhuri, V. N. Smolyaninova, and R. L. Greene, Phys. Rev. **B54,** 14926 (1996)

9. J. M. D. Coey, M. Viret, L. Ranno, K. Ounadjela, Phys. Rev. Lett. **75,** 3910 (1995)

10. E. O. Wollan, W. C. Koehler, Phys. Rev. **100,** 545 (1955) ; *For a recent review regarding this aspect in manganite materials see:* E. L. Nagaev, Physics-Uspekhi **39,** 781 (1996) and references cited therein.

11. Z. Jirak, J. Hejtmanek, E. Pollert, M. Marysko, M. Dlouha, S. Vratislav, J. Appl. Phys. **81,** 5790 (1997)

12. P. D. Battle, M. A. Green, N. S. Laskey, J. E. Millburn, P. G. Radaelli, M. J. Rosseinsky, S. P. Sullivan, J. F. Vente, Phys. Rev. **B54,** 15967 (1996)

13. T. G. Perring, G. Aeppli, Y. Moritomo, and Y. Tokura, Phys. Rev. Lett., **78,** 3198 (1997)

# Figure Captions

Figure 1: X-ray diffraction patterns of $La_{2.3-x}Y_xCa_{.7}Mn_2O_7$ samples with x=0, 0.3, 0.5.

Figure 2: SEM photograph of the $La_{1.8}Y_{.5}Ca_{.7}Mn_2O_7$ sample showing grains and grain boundaries.

Figure 3: (a) Resistivity (ρ) of $La_{2.3-x}Y_xCa_{.7}Mn_2O_7$ x=0, 0.3, 0.5 as a function of temperature. (b) MR as a function of field of $La_{1.8}Y_{.5}Ca_{.7}Mn_2O_7$ at T=4.2 K, 60 K and 100 K.

Figure 4: (a)Magnetization (M) of $La_{2.3-x}Y_xCa_{.7}Mn_2O_7$ x=0, 0.3, 0.5 as a function of temperature measured at 1500 Oe; *(inset)* measured at 84 Oe. (b) Magnetisation vs field for the sample $La_{1.8}Y_{0.5}Ca_{0.7}Mn_2O_7$ at 4.5 K up to 4 Tesla.

Figure 5: Temperature variation of specific heat (C) of $La_{2.3-x}Y_xCa_{.7}Mn_2O_7$ x=0, 0.3, 0.5 up to 24 K. (inset) C/T versus $T^2$ showing linear behavior up to 10 K.

Figure 6: DOS in 2-D and 3-D electronic systems neglecting Coulomb interaction. The Fermi energy ($E_F$) and the DOS at Fermi level ($N(E_F)$) for 65% filling of the band are listed below : (…) for 3-D with $t_x=t_y=t_z=1$, $E_F$ =0.9913 and $N(E_F)$= 0.1405 (—); for 2-D with $t_x=t_y=1$ $t_z=0$, $E_F$=0.6313 and $N(E_F)$=0.1631; (---) 2-D with $t_x$=0.8 $t_y$=0.4 $t_z$=0, $E_F$=.559 and $N(E_F)$=0.244. The area under each is normalized to one.

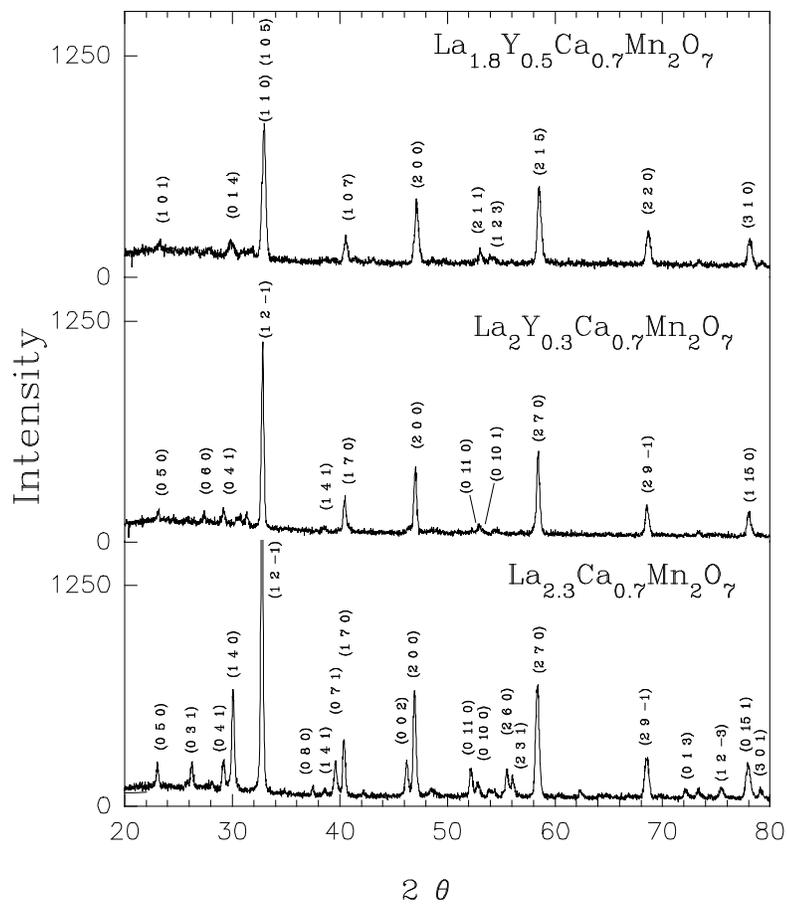

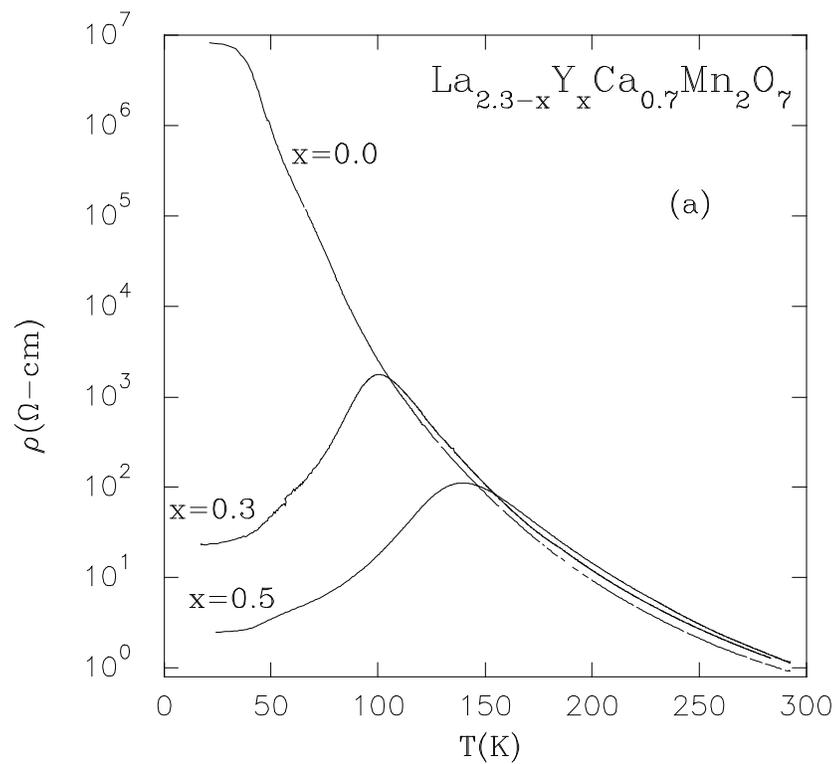

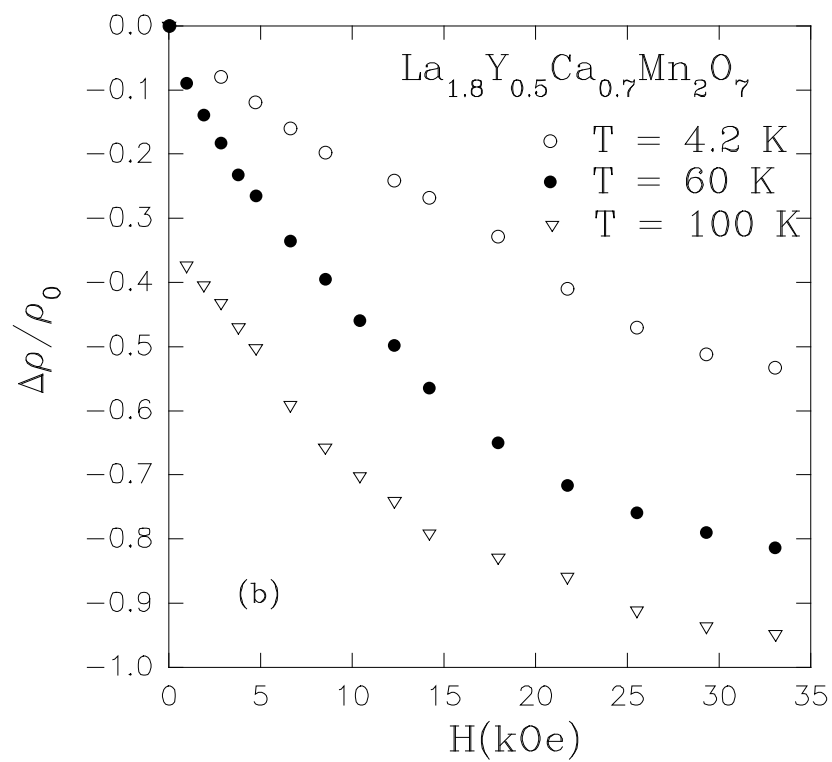

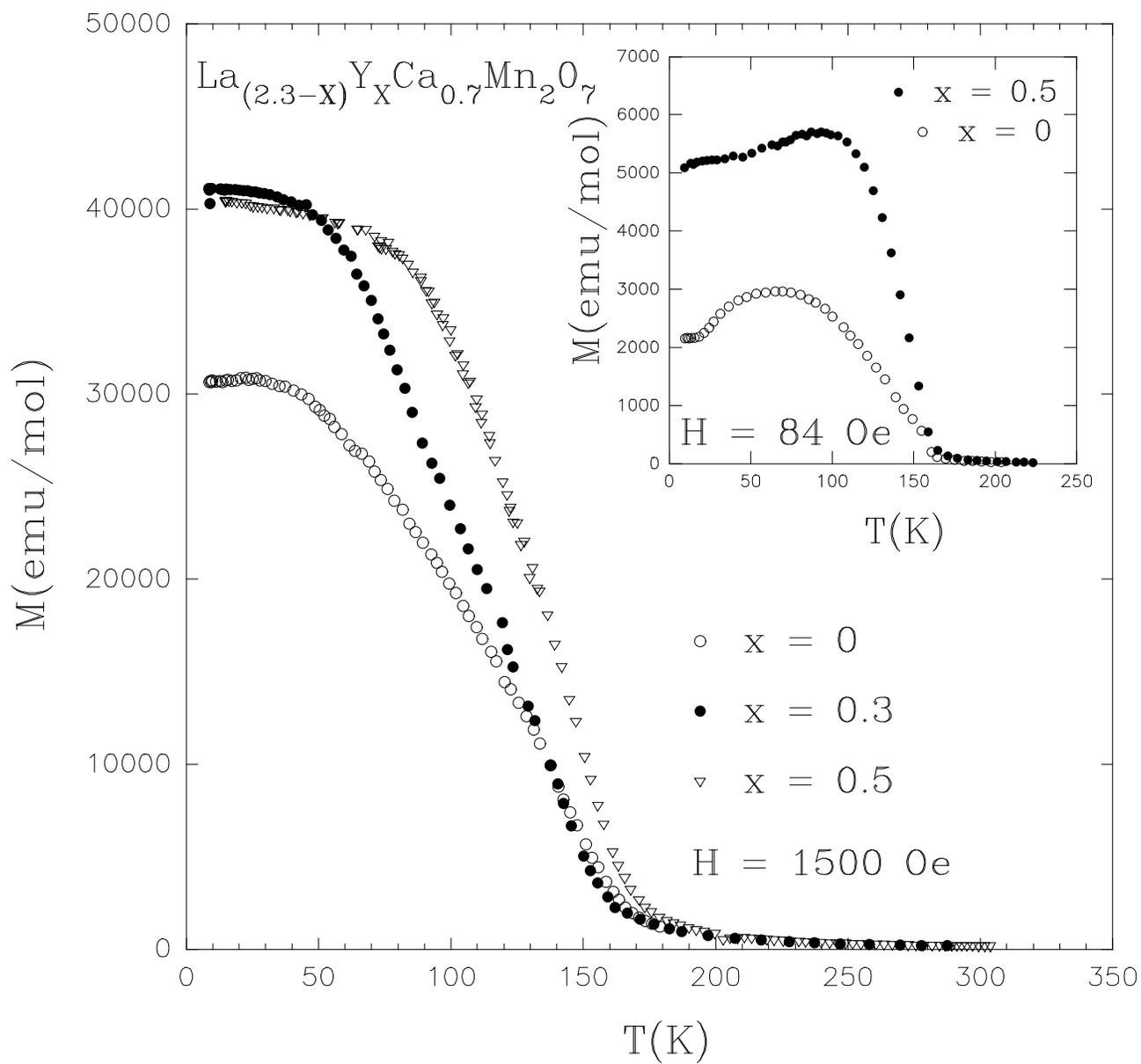

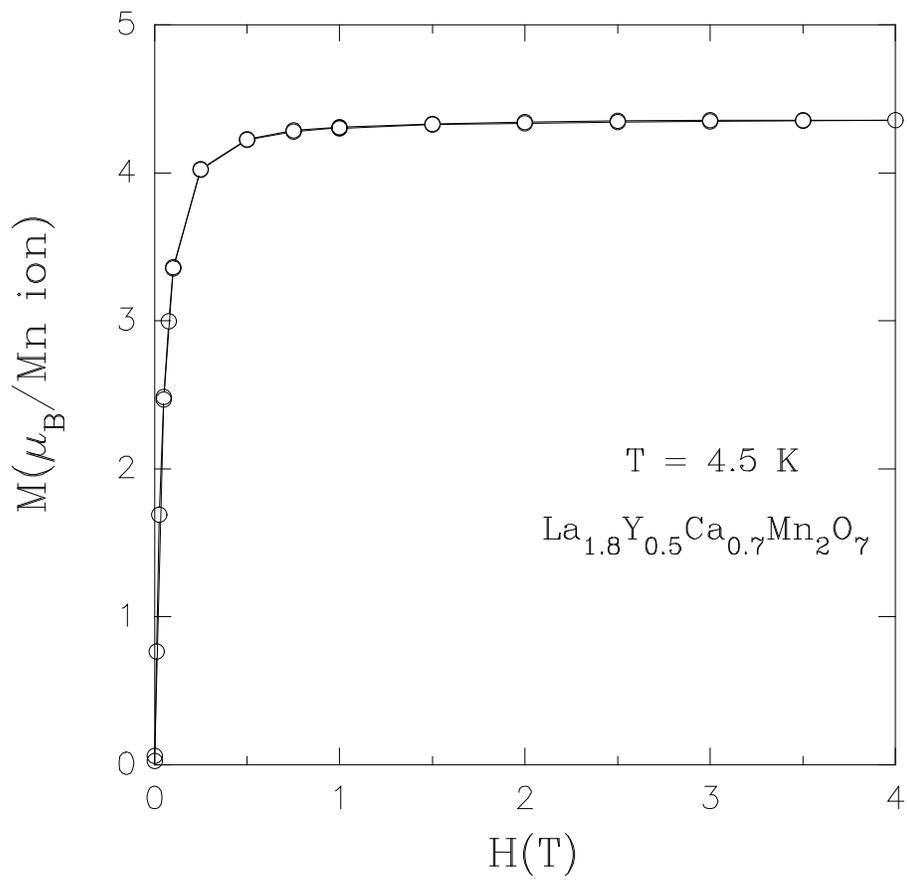

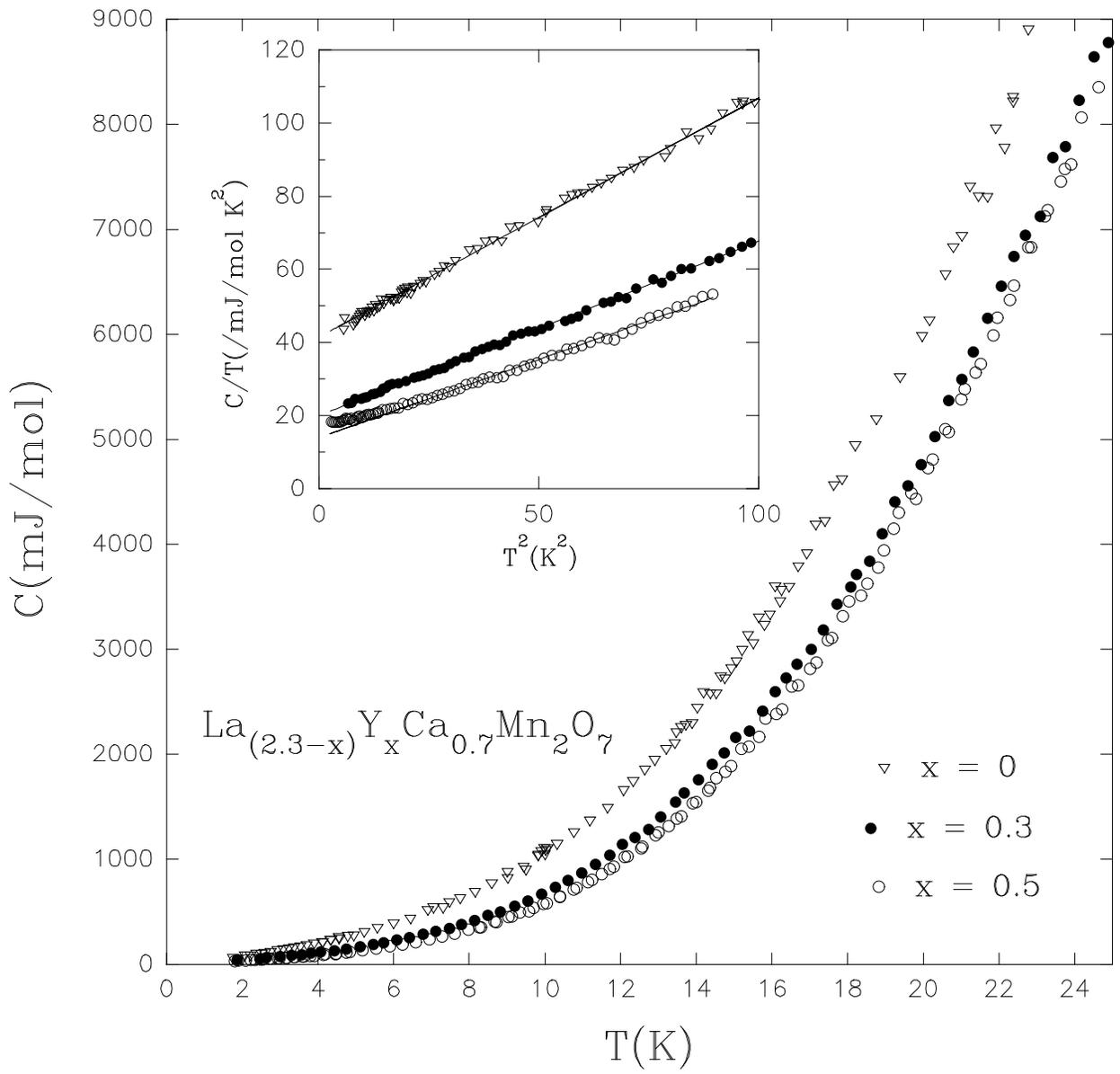

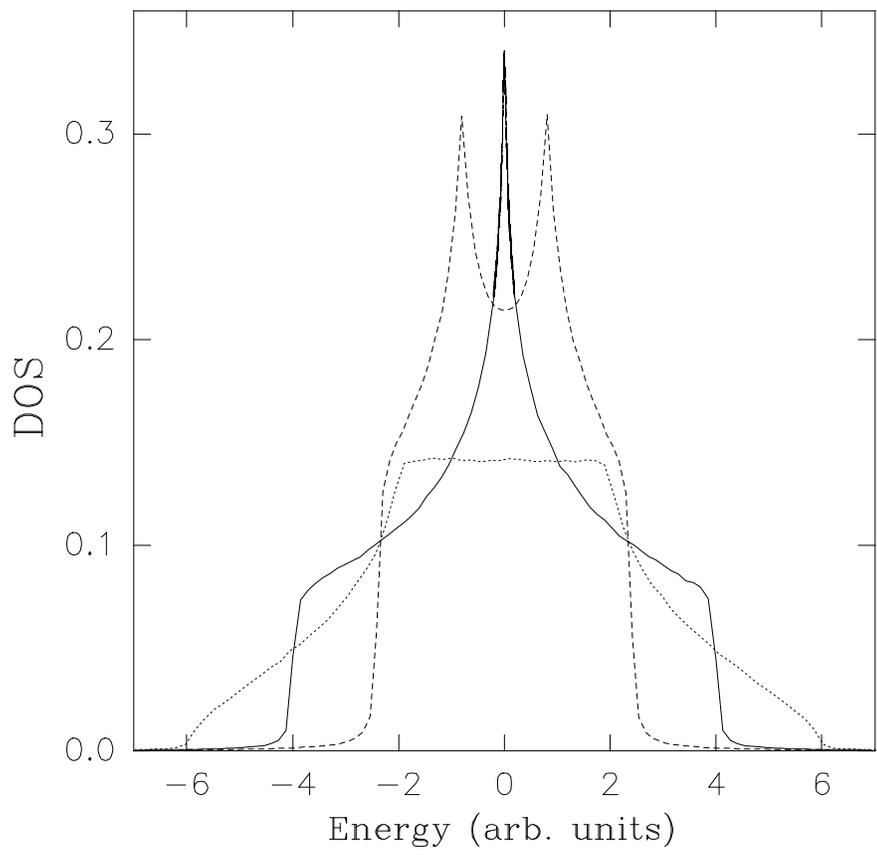